\documentstyle[prc,aps]{revtex}

\begin{document}
\def\Journal#1#2#3#4{{#1} {\bf #2}, #3 (#4)}
\def\NPA{{\em Nucl. Phys. }  A}
\def\PRC{{\em Phys. Rev. }  C}

\draft

\title{
Correlations and realistic interactions in doubly closed shell nuclei 
}
\author{A. Fabrocini$^{1)}$, F.Arias de Saavedra$^{2)}$ and G.Co'$^{\,3)}$}  
\address{
$^1)$ Dipartimento di Fisica, Universit\`a di Pisa,\\ 
and Istituto Nazionale di Fisica Nucleare, sezione di Pisa,\\
I-56100 Pisa, Italy \protect\\
$^2)$ Departamento de Fisica Moderna, 
Universidad de Granada, \\
E-18071 Granada, Spain \protect\\
$^3)$
Dipartimento di Fisica, Universit\`a di Lecce \\
 and Istituto Nazionale di
Fisica Nucleare, sezione di Lecce,  \\  
I-73100 Lecce, Italy }
\maketitle

\date{\today}

\begin{abstract} 

We review the latest variational calculations of 
the ground state properties of doubly closed shell nuclei, 
from $^{12}$C to $^{208}$Pb, with semirealistic and realistic 
two-- and three--nucleon interactions. The studies are carried on 
within the framework of the correlated basis function 
theory and integral equations technique, with state 
dependent correlations having central and tensor components. 
We report results for the ground state energy, one-- and 
two--body densities and static structure functions. For $^{16}$O 
and $^{40}$Ca we use modern interactions and find that the accuracy 
of the method is comparable to that attained in nuclear matter with 
similar hamiltonians, giving nuclei underbound by $\sim$2 MeV/A. The 
computed Coulomb sums are in complete agreement with the latest 
analysis of the experimental data.

\end{abstract} 

\section{Introduction: interaction and correlations}
 Our knowledge of the nuclear interaction is steadily improving 
because of both the ever increasing number of experimental data 
and the more and more sophisticated theories which are being 
developed to tackle the longstanding problem of an accurate 
description of the strongly interacting nuclear systems. The 
interplay between experiments and theory is clear: interactions 
are built which fit the data (typically from nucleon--nucleon 
scattering) and are then tested theoretically in more complicated 
structures. This approach has led to the construction of nucleon--nucleon 
(NN) potentials which reproduce a huge amount of NN scattering data 
(the latest versions fit $\sim$1800 $pp$ and $\sim$2500 $np$ data 
with $\chi^2\sim$1 and break the charge independence and charge 
symmetry \cite{A18,Nijmegen,CD Bonn}). 
However, the use of these accurate, modern NN potentials in 
A$>$2 nuclei has also proved their inability to correctly describe the 
nuclear binding. Light nuclei are underbound and the nuclear matter 
saturation density is not correctly reproduced by a hamiltonian 
containing two--nucleon forces only. Several cures to these 
pathologies have been proposed, ranging from relativistic effects to 
extra degrees of freedom ($\Delta$'s) or to many--body forces. 

If we choose to advocate the lack of binding to the presence of 
more--nucleon potentials, then the first step consists in  the 
introduction of three--nucleon interactions (TNI). The knowledge 
of TNI is by far less deep than that of the NN potential because of 
the considerably lower number of experimental data and of 
the higher difficulty 
in constructing theoretical models. However, even with these 
caveats, the present models of TNI provide the correct binding for 
A=3,4 nuclei (actually, TNI contain parameters which are fitted on 
light nuclei binding energy) and, more important and less obvious, 
bring the computed nuclear matter saturation density very close 
to its empirical value of $\rho_{NM}=$0.16 fm$^{-3}$. 

 Following these guidelines, one can write a realistic nuclear hamiltonian 
in the form:
\begin{equation}
H=-\frac{\hbar^2}{2m} \sum_i \nabla^2_i+\sum_{i<j}v_{ij}+\sum_{i<j<k}v_{ijk} 
\label{eq:hamiltonian}
\end{equation}
where $v_{ij}$ and $v_{ijk}$ are two-- and three--nucleon potentials. The 
NN potential may be split into:
\begin{equation}
v_{ij}=v^{em}_{ij}+v^{\pi}_{ij}+v^{R}_{ij} 
\label{eq:v2}
\end{equation}
where the electromagnetic part, $v^{em}_{ij}$ contains Coulomb, Darwin--Foldy, 
vacuum polarization and magnetic moments contributions; 
the long range part  is well established from a microscopic point of view 
as being given by the One Pion Exchange potential (OPEP), 
$v^{\pi}_{ij}$; the intermediate and short range potential, $v^{R}_{ij}$, 
arises from the exchange of more pions and heavier mesons 
($\rho$ and $\omega$), 
however its microscopic derivation from meson exchange theory is 
troublesome (for the larger number of Feynman diagrams involved  and 
the greater difficulty in the potential extraction) and a phenomenological 
approach is often followed, consisting in a physically plausible 
parametrization in terms of a reasonable number of parameters. For instance, 
the Argonne $v_{18}$ NN potential \cite{A18} 
 contains 43 parameters and it is given by a sum of 18 terms
\begin{equation}
 v_{ij} =\sum_{p=1,18}v_p(r_{ij})O^p_{ij}~~, 
\label{eq:v18}
\end{equation}
where the first 14 terms are isoscalar, with 
\begin{equation}
 O^{p=1,14}_{ij}=\left[1,{\bf\sigma}_i\cdot{\bf\sigma}_j,
S_{ij},{\bf L}\cdot{\bf S}, {\bf L}^2,{\bf L}^2{\bf\sigma}_i\cdot{\bf\sigma}_j,
({\bf L}\cdot{\bf S})^2\right]\otimes
\left[1, {\bf\tau}_i\cdot{\bf\tau}_j\right]~~,  
\label{eq:op14}
\end{equation}
and the remanining isovector and isotensor components are:
\begin{equation}
 O^{p=15,18}_{ij}=\left[1,{\bf\sigma}_i\cdot{\bf\sigma}_j,S_{ij}\right]
\left[3\tau_{iz}\tau_{jz}- {\bf \tau}_i\cdot{\bf \tau}_j\right],
\left[\tau_{iz}+\tau_{jz}\right]~~.
\label{eq:op18}
\end{equation}

Two--pion exchange gives large part of the attraction of the three--nucleon 
potential, whereas the remainder is mostly phenomenological. 
Recently, attempts are being made to evaluate microscopically other 
pieces, as the three--pion two--Delta diagram \cite{Pieper}. 
A popular form of TNI are the Urbana forces \cite{UrbanaTNI}, 
given by
\begin{equation}
v_{ijk}=v^{2\pi}_{ijk}+v^{R}_{ijk}~~. 
\label{eq:v3}
\end{equation}
The two--pion Fujita--Miyazawa \cite{Fujita} term, 
$v^{2\pi}_{ijk}$, is
\begin{equation}
v^{2\pi}_{ijk}=\sum_{cycl}
A_{2\pi}\{X_{ij},X{ik}\}
\{{\bf\tau}_i\cdot{\bf\tau}_j,{\bf\tau}_i\cdot{\bf\tau}_k\}+
C_{2\pi}\left[X_{ij},X{ik}\right]
\left[{\bf\tau}_i\cdot{\bf\tau}_j,{\bf\tau}_i\cdot{\bf\tau}_k\right]~~,  
\label{eq:v2pi}
\end{equation}
with
\begin{equation}
X_{ij}=Y_\pi(r_{ij}){\bf \sigma}_i\cdot{\bf \sigma}_j+
T_\pi(r_{ij})S_{ij}~~,
\end{equation}
$Y_\pi(r)$ and $T_\pi(r)$ being the Yukawa and tensor Yukawa functions. The
short range part, $v^{R}_{ijk}$, is written as
\begin{equation}
v^{2\pi}_{ijk}=\sum_{cycl}
U_0 T^2_\pi(r_{ij})T^2_\pi(r_{ik})~~. 
\label{eq:v3R}
\end{equation}
$A_{2\pi}, C_{2\pi}$ e $U_0$ are chosen to reproduce the binding of 
 A=3,4 nuclei.

 Once a realistic hamiltonian has been built, task of the theorist is 
to address the ground state Schroedinger equation, $H\Psi_0=E_0\Psi_0$, 
to solve it or to find good approximations to its solutions. 
The exact solution is now possible in light nuclei by a variety of techniques: 
Green's Function Monte Carlo \cite{GFMC} for 3$\leq$A$\leq$8, 
Faddeev and Faddeev--Yakubovsky \cite{Faddeev} and Correlated Hyperspherical 
expansions (CHE) \cite{CHE}.
GFMC has been pushed to the highest A--value, whereas the other approaches 
have been used mainly for A=3,4. However, they are more flexible in the 
sense that both Faddeev and Correlated Hyperspherical expansion methods 
are more easibly extendible to the study of three-- and four--body scattering 
reactions. 
In fact, several few--body reactions of relevant astrophysical interest 
have been accurately analyzed by CHE. For the remaining part of the 
nuclear table, including nuclear matter, methods to exactly solve the 
Schroedinger equation are still to come, even if a very promising approach,  
based on the Hubbard--Stratonovich transformation, is currently under 
development \cite{Hubbard}. 
Hence, alternative ways to gather information on the 
many--body wave function must be devised. The Brueckner theory has brought 
standard perturbative theories to an extremely high level of 
sophistication \cite{Brueckner}; the Coupled Cluster (CC) method has been 
used to build correlations into the wave function and is, by now, a standard 
technique in many--body physics \cite{coupled}; the variational principle has 
provided a powerful recipe  to construct accurate wave functions.

A direct consequence of the strong nuclear interaction is the failure 
of the mean field (MF) approach in the  description of the nucleus. A 
striking evidence of this fact is provided by $(e,e')p$ experiments 
\cite{ee'p} where a clear signature of a depletion of the occupation 
probability with respect to unity for states below the Fermi level,
 $\epsilon_F$, (as requested by MF theories) is present, together with a 
non zero occupation probability for states above $\epsilon_F$. 
This behavior may be explained in terms of short range 
{\it dynamical correlations} (in contrast with {\it statistical correlations}, 
due to the antisymmetry) generated in the wave functions by the interaction and 
which can be hardly described by standard perturbation methods based on 
a non interacting basis. For instance, Brueckner theory must sum infinite 
numbers of ladder diagrams in order to deal efficiently with the nuclear 
potential.

In the Correlated Basis Function (CBF) theory the non perturbative 
correlation effects are directly embedded into the basis functions. 
This property makes the theory a powerful tool to investigate 
many-body interacting systems in several fields of physics, 
as liquid Helium, electronic structures (both in the forms of electron 
fluids and lattices) and both finite nuclei and infinite nuclear matter. 
The flexibility of the CBF approach results in a realistic description 
not only of the ground state (energy, momentum distribution, 
distribution functions and so on) but also of dynamical (cross sections) 
quantities. 
 
A set of correlated basis wave functions, $\Psi_n(1,2...A)$, 
 may be built by applying a many--body correlation operator, 
$F(1,2...A)$, to the model basis functions, $\Phi_n(1,2...A)$, 
\begin{equation}
\Psi_n(1,2...A)=F(1,2...A)\Phi_n(1,2...A)~~,
\label{eq:Psi_n}
\end{equation}
where the operator $F$ is intended to take care of the dynamical 
correlations,  whereas the model wave functions, $\Phi_n$, include 
antisymmetrization effects and, possibly, long range correlations 
due to collective excitations (as BCS type states or surface effects). 
A perturbative theory may ve developed in terms of the correlated 
states (\ref{eq:Psi_n}), having the nice property of a rapid convergence 
and the counterindication of a large difficulty in computing the 
matrix elements. The zeroth order of this 
theory is often referred to as the {\it variational}  level of CBF. In fact, 
the variational principle is used to fix the correlation operator by 
minimizing the ground state energy, 
$E_0^v=\langle \Psi_0|H|\Psi_0\rangle /\langle \Psi_0|\Psi_0\rangle$. 

The choice of the correlation operator depends, to a large extent, on 
the interaction. A form of $F(1,2...A)$ that has shown to be suitable 
to nuclear systems is 
\begin{equation}
F(1,2...A)= {\cal S}\left\{\prod_{i<j=1,A}F_{ij}\right\}~~,
\label{eq:F}
\end{equation}
{\it i.e.} a symmetrized product of two-body correlation operators, $F_{ij}$.  
The model states,  $\Phi_n(1,2...A)$, are Slater determinants of 
single particle wave functions, $\phi_\alpha(i)$, obtained by some MF 
potential (simple plane waves in infinite, homogeneous matter).
 $F_{ij}$ is chosen, consistently with the interaction, to be given by 
\begin{equation}
F_{ij}=\sum_{p=1,8}f^p(r_{ij})O^p_{ij}~~,
\label{eq:F_ij}
\end{equation}
where the sum runs up to the spin--orbit components in eq.(\ref{eq:op14}). 
If the $p\geq 2$ components are disregarded, the Jastrow scalar correlation 
is recovered.
The free minimization of $E_0^v$ would provide the best choice for the 
correlation functions $f^p(r)$. 
However, this is a prohibitive task in nuclear systems, so 
the correlation functions are usually parametrized and the 
parameters are fixed by minimizing $E_0^v$. An additional minimization 
may be performed on the parameters of the MF model wave function.

\section{Something on cluster expansions and Fermi Hypernetted Chain equations}

 As already mentioned in the previous Section, computing the matrix 
elements in the CBF approach represents the greatest difficulty in 
its application. In principle one could use Monte Carlo (MC) based 
algorithms to sample the occuring multi-dimensional integrals. 
For simple interactions and correlations this can be done even for 
very large numbers of particles. In such a case, the $A=\infty$ limit 
can also be studied by means of appropriate conditions at the borders 
of the simulation box. Liquid Helium, both as a fluid and as droplets, 
has been subject of MC investigations. However, the strong state 
dependence of the nuclear interaction prevents from the use of MC 
techniques in medium--heavy nuclei, as well as in nuclear matter.
The $^{16}$O ground state was studied within CBF by 
the cluster Monte Carlo (CMC) method \cite{CMC}, where
the Jastrow correlations contribution is exactly treated by MC 
sampling, and  that from the remaining operatorial components is 
approximated by considering (via MC) up to four-- or five--body 
cluster terms. 

An alternative approach is provided by cluster expansions. 
The expectation values of operators are written in terms of 
$n$--body densities, 
\begin{equation}
 \rho_1(1)=\langle \sum_i \delta ({\bf r}_i-{\bf r}_1)\rangle~~, 
\label{eq:rho_1}
\end{equation}
\begin{equation}
 \rho_2^{(p)}(1,2)=\langle \sum_{i\neq j} 
\delta ({\bf r}_i-{\bf r}_1)\delta ({\bf r}_j-{\bf r}_2)O^p_{ij} 
\rangle~~,
\label{eq:rho_2}
\end{equation}
and related quantities, as the density matrices. The densities are then 
cluster expanded in terms of dynamical, $h(r)=[f^1(r)]^2-1$ and 
$f^{1}(r)f^{p\geq 2}(r)$, and statistical, 
$\rho_0(i,j)=\sum_\alpha \phi_\alpha^\dagger (i)\phi_\alpha (j)$,
 correlations. 
The expansion is {\sl linked}, in the sense that disconnected diagrams,  
given by the product of two or more pieces not connected among each 
other by any kind of correlation and coming from the expansion of the 
numerator, cancel exactly with those generated by the expansion of 
the denominator. However, the cluster expansion is not {\it irreducible}, 
{\it i.e.} the surviving connected diagrams bear vertex corrections 
 (except for the Jastrow correlation in the A=$\infty$ case, where the 
vertex correction is simply the density). 

 The cluster diagrams are then classified into: {\sl Nodal}, N, 
where all the paths between the external points (for instance, points 
1 and 2 for the two--body density of eq.(\ref{eq:rho_2})) go through 
the same internal point, or {\sl node}; {\sl Composite}, C, 
given by the product of two or more nodals; the remaining 
diagrams belong to the {\sl Elementary} class, E.

Fermi Hypernetted Chain (FHNC) integral equations \cite{FHNC} 
allow for the exact summation of the nodal and composite diagrams, 
in the Jastrow case, once the sum of the elementary diagrams, 
$E(1,2)$, is given. So, the function $E(1,2)$ is actually an 
input for solving thr FHNC equations. Unfortunately, no exact 
way of computing this function is known and one must use 
approximations. The simplest one is the FHNC/0 truncation, 
where the choice $E=0$ is made. This seemingly crude approximation 
is however appropriate to nuclear systems, where the density is not 
very high. In denser liquid Helium, a more accurate treatment of 
the elementary diagrams is needed.   

For state dependent correlations, a complete FHNC summation is possible 
only for the Jastrow part. In fact, different orderings of the 
operators in the symmetrized product (\ref{eq:F}) may give the 
same cluster contribution and a scheme to correctly keep track of 
all of them has not been devised so far.  However, partial classes 
of diagrams containing operatorial correlations may be exactly 
summed by the Single Operator Chain approximation (FHNC/SOC) \cite{SOC}. The 
FHNC/SOC integral equations sum all the nodal diagrams with only 
one operatorial correlation per internal side, besides all the Jastrow 
correlated clusters. The accuracy of the FHNC/SOC approximation 
has been set to $\sim$1 MeV/A in nuclear matter at saturation 
density by computing its leading corrections \cite{MOC}. The 
FHNC/SOC scheme has been recently extended to doubly closed 
shell nuclei in $ls$ coupling \cite{SOC_nuclei}. 

A way to control the approximations is provided by the 
density sum rules. In fact, the exact densities satisfy the 
following normalizations:
\begin{equation}
 \int d{\bf r}_1\rho_1(1)=A ~~,
\label{eq:SR1}
\end{equation}
\begin{equation} \frac{1}{A(A-1)}
 \int d{\bf r}_1\int d{\bf r}_2 \rho_2^{(1)}(1,2)=~1 ~~,
\label{eq:SR2}
\end{equation}
\begin{equation} \frac{1}{3A}
 \int d{\bf r}_1\int d{\bf r}_2 \rho_2^{(\tau)}(1,2)=-1 ~~.
\label{eq:SRtau}
\end{equation}
The last equality holds for isospin saturated systems. Deviations from the 
sum rules are a powerful check for the approximations made in the 
solution of the FHNC equations, both in the treatment of the elementary 
diagrams and of the operatorial correlations.

As a matter of fact, FHNC/SOC satisfies the normalizations of the 
densities to a high degree of accuracy, as it is shown in
Table~\ref{tab:sumrules}. The Table gives the one--body 
(S$_1$, eq.~(\ref{eq:SR1}))and  central and isospin two--body
(S$_2$ and S$_\tau$, eqs.~(\ref{eq:SR2},\ref{eq:SRtau}) density 
normalizations for  Jastrow ($f_1$) and operatorial ($f_6$) 
correlations in $^{16}$O and $^{40}$Ca. 
The $f_6$ model does not contain spin--orbit 
components but it does have the tensor ones. The Jastrow model 
satisfies practically exactly the sum rules, whereas the SOC 
approximation produces discrepancies of 
$\sim$9\% in the worst case \cite{SOC_nuclei}.

\section{Energies, densities and structure functions}

 In a series of papers \cite{nuclei1,nuclei2,nuclei3} we extended 
the FHNC formalism to doubly closed shell nuclei, both in $ls$ 
and $jj$ coupling scheme, described by Jastrow correlated wave 
functions. In those papers model interactions were used, as the 
semirealistics Afnan and Tang (S3) one \cite{S3}.  The original 
S3 potential is purely central, without tensor components, 
and reproduces the NN $s$--waves scattering data up to $\sim$60 MeV. 
Its modified version \cite{MS3} has been supplemented in the odd 
channels of the same repulsion as in the even ones and gives a 
nuclear matter binding curve close to the current realistic potentials. 
The correlations we adopted contained a $\tau_z$ dependence, allowing 
for distinguishing between the different nucleon pairs 
($f_{nn}\neq f_{pp}\neq f_{np}$). The ground state energies for the 
$\tau_z$--dependent correlations are shown in Table~\ref{tab:nuclei_jj} 
together with the experimental values. The correlations have been 
fixed by solving the Euler equations which result from the minimization of
the expectation value of the hamiltonian at the second order of the 
cluster expansion. 
Given the poor quality of the interaction, the 
comparison with the experiment is not really meaningful, even if not 
completely unsatisfactory. However, it is interesting to note a saturation 
trend of the binding energy along the mass number. The Table also gives the 
energies obtained by an isospin independent correlation, identical for all 
the pairs (Average Correlation Approximation, ACA). The poorer variational 
quality of the ACA correlation produces higher energies than the isospin 
dependent model. The density normalizations are always satisfied with 
an accuracy greater than percent for the calculations of the Table.
 
ooooooooo

FHNC/SOC has been pushed to deal with 
realistic potentials and state dependent, non central correlations 
in $^{16}$O and $^{40}$Ca in Refs.\cite{SOC_nuclei,TNI_nuclei}. 
 In the last work a complete hamiltonian of the form given 
in eq.(\ref{eq:hamiltonian}) 
with the inclusion of the momentum dependent parts of the NN 
interaction and of the three--nucleon potential, as well as of the Coulomb 
term, was adopted. Table~\ref{tab:nuclei_A14} gives the 
g.s. energies  per nucleon for two models: 
 Argonne $v_{14}$ + Urbana VII \cite{A14,UrbanaVII} (A14+UVII) 
and Argonne $v'_{8}$ + Urbana IX \cite{GFMC} (A8'+UIX). 
The correlated wave 
function features a $f_6$ nuclear matter correlation 
and single particle wave functions generated by 
a Woods--Saxon mean field potential. 
 In the A14+UVII case, the $^{16}$O FHNC/SOC energies are compared with 
the CMC results of Ref.\cite{CMC}, obtained  with a 
similar wave function, and with the Coupled Cluster ones of Ref.\cite{CCM}. 
The r.m.s. radius is also given. The variational A14+UVII energies are 
close to the CC estimates, while the discrepancies with CMC are due to 
the approximations in the FHNC/SOC scheme and to small differencies 
in the correlation. Actually, 
our nuclear matter Euler equation does not contain the spin--orbit 
potential, whereas that used for the CMC correlations does. 
This fact produces small variations, which mostly affect the 
kinetic energy. FHNC and CMC give similar radii, both of them 
smaller than CC. Explicit three--body correlations, 
not given by the product of two--body ones, were found to provide 
$\sim$0.85 MeV/A extra binding \cite{CMC} in the A14+UVII case.
 
 The charge densities, $\rho_c(r_1)$, and the two--body distribution 
functions, $\rho_2(r_{12})$, computed with the A8'+UIX 
wave function are shown in Figure \ref{fig:densities}. The CBF $\rho_c(r_1)$ 
(dashed lines) are folded with the proton form factor and compared with 
the experimental data (solid lines). $\rho_2(r_{12})$ is defined as    
\begin{equation}
\rho_2( r_{12})=\frac {1}{A}\int d^3 R_{12}
\rho_2^{(1)}({\bf r}_1,{\bf r}_2)~~,
\label{2NDF}
\end{equation}
where ${\bf R}_{12}$ is the center of mass coordinate. 
Short range correlations strongly affect $\rho_{2}(r_{12})$ (solid lines)  
at small internucleon distances, where the NN repulsion heavily depletes 
the distribution functions with respect to the IPM estimates (dashed lines).  

Two--body densities are also of great interest since they allow 
for analyzing several integrated nuclear cross sections. 
In fact, the responses of a nucleus to an external probe,  
described by an operator $O_X$,  
can be expressed in terms of the dynamical structure 
functions, $S_X(q,\omega)$, whose 
non energy weighted sum gives the static structure function (SSF), 
$S_X(q)$. In turn, the SSF are given by expectation 
values of appropriate combinations of one-- and two--body densities. 
For instance, the electromagnetic longitudinal response, $S_L(q,\omega)$, 
as  measured in electron--nucleus experiments, is mostly due 
to charge fluctuations. The corresponding SSF, $S_L(q)$ 
(or Coulomb sum), is given by 
\begin{equation}
S_L(q)= 1 + {1\over {Z}} \int d^3r_1 \int d^3r_2
~~e^{\imath {\bf q}\cdot {\bf r}_{12} }
\left[\rho_{pp}({\bf r}_1,{\bf r}_2)-
\rho_c(r_1) \rho_c(r_2)\right]~~,
\label{Sc_q}
\end{equation}
where $\rho_{pp}({\bf r}_1,{\bf r}_2)$ is the 
proton--proton two--body density. 

In Figure \ref{fig:coulomb_sum} we compare the A8'+UIX 
CBF Coulomb sums with those extracted from the 
world data on inclusive quasi--elastic electron scattering \cite{jou96} 
experiments in $^{12}$C, $^{40}$Ca, and $^{56}$Fe. 
The Figure also shows the nuclear matter Coulomb sum for the 
A14+UVII model from Ref. \cite{sch87}. The agreement of the 
nuclei theoretical SSF with the latest experimental data, 
which properly take into account the large energy tail of 
the response, is complete, whereas  
the nuclear matter shows some discrepancies at the 
lowest $q$ values where finite size effects appear to be still relevant.

\section{Conclusions}

A noticeable progress in the microscopic study of the ground state 
structure of medium--heavy, doubly closed shell nuclei has been made 
in the latest years. Going beyond the simple mean field picture and 
using realistic hamiltonians seems now to be within reach of 
many--body theories. In this context, the variational approach 
 and the correlated basis function theory are among the most 
promising tools. Fermi hypernetted chain and single operator chain 
techniques appear to deal with the complexity of the interaction and 
of the correlation with the same degree of accuracy that has been 
achieved in nuclear matter. This opens the field both to the microscopic 
study of dynamical quantitites, as the nuclei cross sections, 
and to the extension of the variationl method to other interesting systems,
 as, for instance, hypernuclei.

%
%
\newpage

\begin{table}
\caption{One-- and two--body density sum rules in $^{16}$O and $^{40}$Ca 
for Jastrow and operatorial correlations.}
\begin{tabular}{ccccc}
 & & S$_1$ & S$_2$ & S$_\tau$ \\
\tableline
           &       &       &      &       \\ 
$^{16}$O   & $f_1$ & 16.00 & 1.00 & -1.00 \\ 
           & $f_6$ & 16.01 & 1.05 & -0.94 \\ 
           &       &       &      &       \\ 
\tableline
           &       &       &      &       \\ 
$^{40}$Ca  & $f_1$ & 40.00 & 1.00 & -1.00 \\ 
           & $f_6$ & 39.86 & 1.09 & -0.98 \\ 
           &       &       &      &       \\ 
\end{tabular}
\label{tab:sumrules}
\end{table}

\begin{table}
\caption{Ground state energies for doubly closed shell nuclei with the 
S3 potential and Jastrow correlated wave functions. The EUL and ACA 
 lines give the energies with $\tau_z$ dependent and independent 
correlations, respectively. Energies in MeV.}
\begin{tabular}{ccccc}
 & & $\langle$H$\rangle$ & E/A & (E/A)$_{expt}$ \\
\tableline
           &       &       &      &       \\ 
$^{12}$C   &  EUL  & -33.2 & -3.84& -7.68 \\ 
           &  ACA  & -15.5 & -2.36&       \\ 
           &       &       &      &       \\ 
\tableline
           &       &       &      &       \\ 
$^{16}$O   &  EUL  & -119.7& -8.20& -7.98 \\ 
           &  ACA  & -82.8 & -5.89&       \\ 
           &       &       &      &       \\ 
\tableline
           &       &       &      &       \\ 
$^{40}$Ca  &  EUL  & -381.2& -9.78& -8.55 \\ 
           &  ACA  & -262.1& -6.81&       \\ 
           &       &       &      &       \\ 
\tableline
           &       &       &      &       \\ 
$^{48}$Ca  &  EUL  & -394.9& -8.43& -8.67 \\ 
           &  ACA  & -272.3& -5.87&       \\ 
           &       &       &      &       \\ 
\tableline
           &       &       &      &       \\ 
$^{208}$Pb &  EUL  &-1761.4& -8.50& -7.87 \\ 
           &  ACA  &-1056.3& -5.11&       \\ 
           &       &       &      &       \\ 
\end{tabular}
\label{tab:nuclei_jj}
\end{table}

\begin{table}
\caption{$^{16}$O and $^{40}$Ca ground state energies per nucleon and radii 
for the Argonne $v_{14}$ + Urbana VII and the Argonne $v'_{8}$ + Urbana IX 
models with the FHNC/SOC, Cluster Monte Carlo 
(CMC) and Coupled Cluster (CC) methods.} 
\begin{tabular}{cccc}
  & $^{16}$O & E/A (MeV) & rms (fm) \\
\tableline
          &       &       &          \\ 
 A14+UVII &       &       &          \\ 
          &       &       &          \\ 
 $^{16}$O & FHNC  & -5.97 & 2.44     \\ 
          & CCM   & -6.90 & 2.43     \\ 
          & CC    & -6.10 & 2.86     \\ 
          &       &       &          \\ 
\tableline
          &       &       &          \\ 
 A8'+UIX  &       &       &          \\ 
          &       &       &          \\ 
 $^{16}$O & FHNC  & -5.41 & 2.67     \\ 
          & expt  & -7.98 & 2.73     \\ 
          &       &       &          \\ 
 $^{40}$Ca& FHNC  & -6.64 & 3.39     \\ 
          & expt  & -8.55 & 3.48     \\ 
          &       &       &          \\ 
\end{tabular}
\label{tab:nuclei_A14}
\end{table}
%
%
\newpage

\begin{figure}
\caption{CBF charge densities and two--body distribution 
functions for the A8'+UIX interaction.}
\label{fig:densities}
\end{figure}
 
\begin{figure}
\caption{CBF and experimental Coulomb sums for finite nuclei and 
nuclear matter.} 
\label{fig:coulomb_sum}
\end{figure}

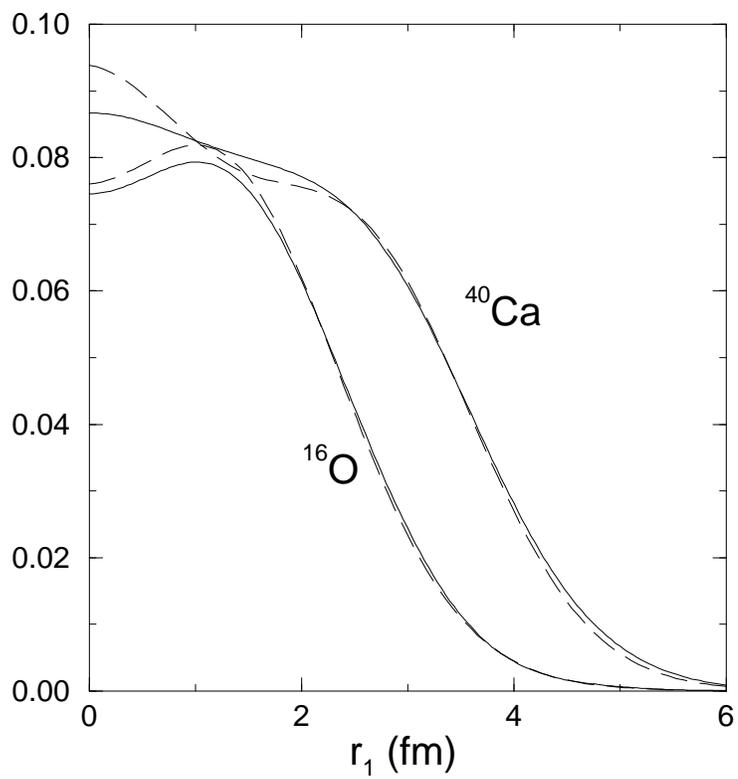
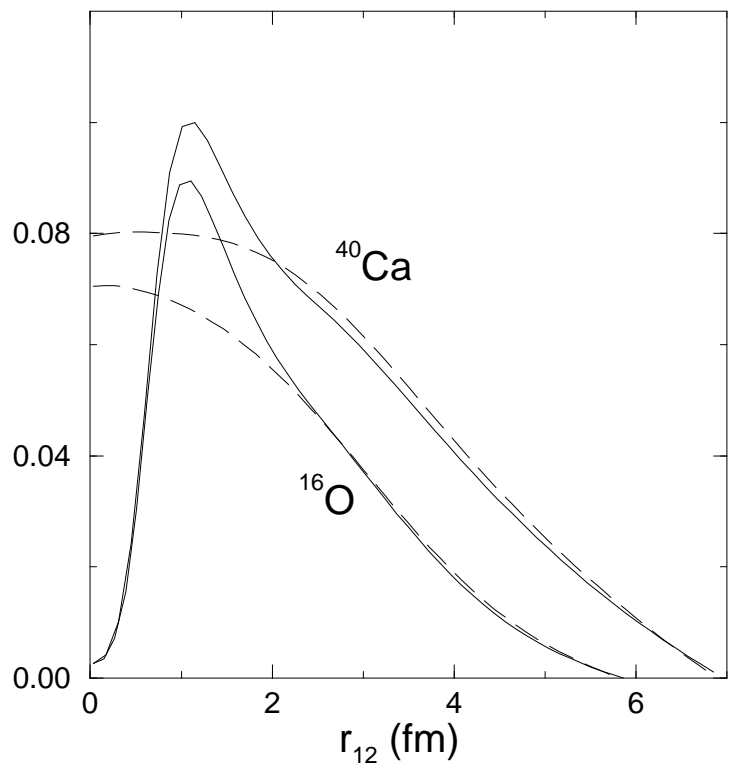
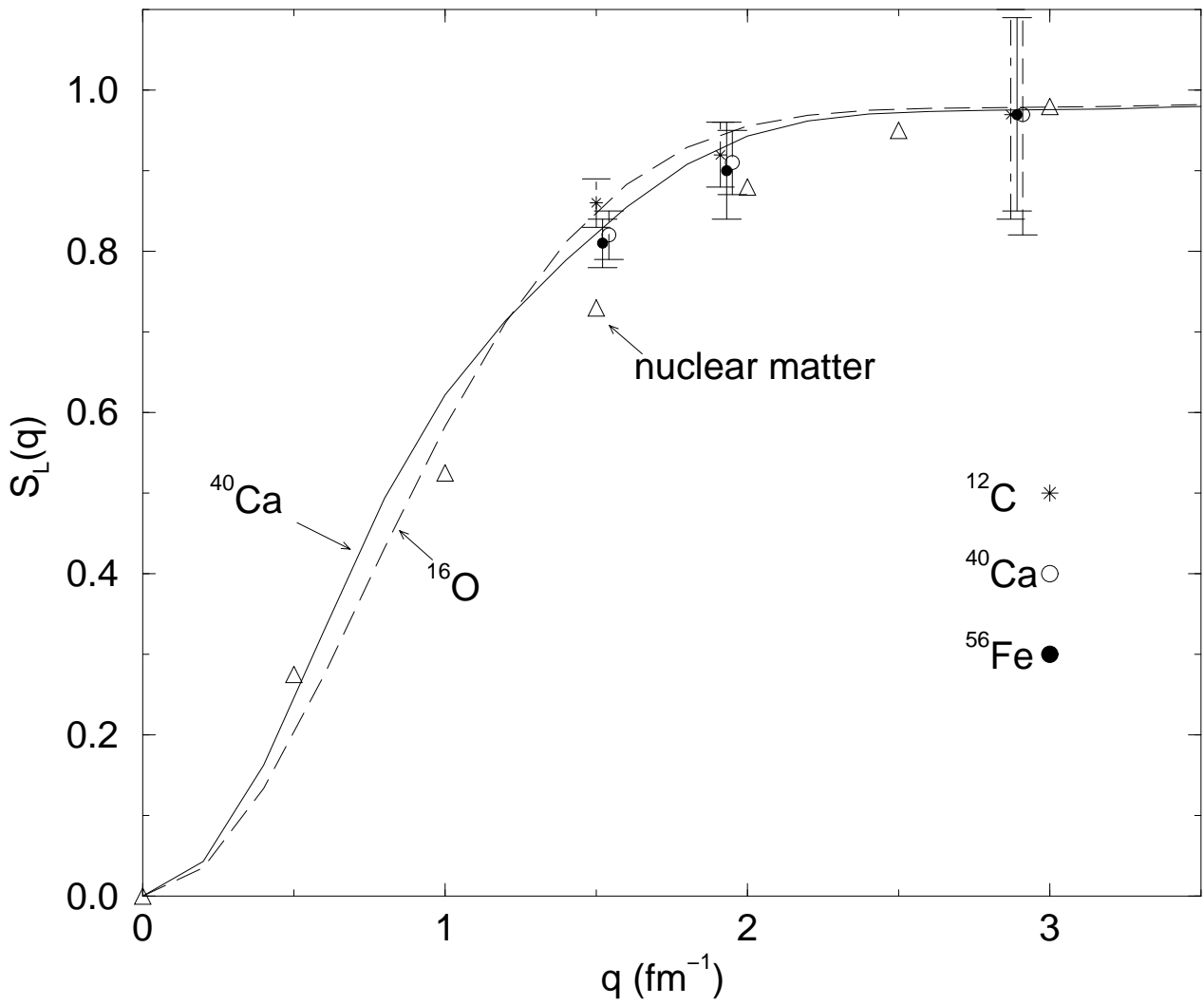
\end{document}